\documentclass[]{eptcs}
\providecommand{\event}{CompMod 2011} 
\usepackage{breakurl}             
\usepackage{epsfig}
\usepackage{amsfonts}
\usepackage{amssymb}
\usepackage{amsmath}
\usepackage[usenames,dvipsnames]{color}
\usepackage{subfigure}
\usepackage{verbatim}

\newif\ifmc
\mcfalse 
\newcommand{\mc}[1]
{\ifmc{\color{Blue}{#1}}\else{#1}\fi}

\newcommand{\es}[1]
{\ifmc{\color{Magenta}{#1}}\else{#1}\fi}

\newcommand{\qqop}[1]{\mathrel{\makebox[2em]{$#1$}}}

\newcommand{\agr}{\quad\big|\quad}

\newcommand{\AT}{\mathcal{A}}
\newcommand{\LT}{\mathcal{L}}

\newcommand{\LeftPat}{p}

\newcommand{\RightPat}{o}

\newcommand{\TOP}{\top}

\makeatletter
\def\mapstofill@{%
\arrowfill@{\mapstochar\relbar}\relbar\rightarrow}
\newcommand*\xmapsto[2][]{%
\ext@arrow 0000\mapstofill@{#1}{#2}}
\makeatother

\newcommand{\srewrites}[1]{\stackrel{#1}{\longmapsto}}

\newcommand{\into}{\ensuremath{\,\rfloor}\,}

\newcommand{\conc}{\;\,}
\newcommand{\emptyseq}{\bullet}

\newcommand{\red}{\longmapsto}

\title{Modelling Spatial Interactions in the Arbuscular Mycorrhizal Symbiosis using the Calculus of Wrapped Compartments\thanks{This research is funded by the BioBITs Project (\emph{Converging
Technologies} 2007, area: Biotechnology-ICT), Regione Piemonte.}}

\author{Cristina Calcagno$^1$, Mario Coppo$^2$, Ferruccio Damiani$^2$, Maurizio Drocco$^2$\\Eva Sciacca$^2$ , Salvatore Spinella$^2$ and Angelo Troina$^2$
\institute{$^1$Dipartimento di Biologia Vegetale, Universit\`a di Torino}
\institute{$^2$Dipartimento di Informatica, Universit\`a di Torino}
}
\def\titlerunning{Modelling Spatial Interactions in the AM Symbiosis using CWC}
\def\authorrunning{C. Calcagno, M. Coppo, F. Damiani, M. Drocco, E. Sciacca, S. Spinella and A. Troina}
\begin{document}
\maketitle

\begin{abstract}

Arbuscular mycorrhiza (AM) is the most wide-spread plant-fungus symbiosis on earth. Investigating this kind of symbiosis is considered one of the most
promising ways to develop methods to nurture plants in more natural manners, avoiding the complex chemical productions used nowadays to produce artificial
fertilizers. In previous work we used the Calculus of Wrapped Compartments (CWC) to investigate different phases of the AM symbiosis. In this paper, we
continue this line of research by modelling the colonisation of the plant root cells by the fungal hyphae spreading in the soil. This study requires the
description of some spatial interaction. Although CWC has no explicit feature modelling a spatial geometry, the compartment labelling feature can be
effectively exploited to define a discrete surface topology outlining the relevant sectors which determine the spatial properties of the system under
consideration. Different situations and interesting spatial properties can be modelled and analysed in such a lightweight framework (which has not an
explicit notion of geometry with coordinates and spatial metrics), thus exploiting the existing CWC simulation tool.
\end{abstract}

\section{Introduction}
\label{intro}

Arbuscular mycorrhiza (AM) is the most wide-spread plant-fungus symbiosis on earth~\cite{harrison2005signaling}. Investigating this kind of symbiosis is
considered one of the most promising ways to develop methods to nurture plants in more natural manners, avoiding the complex chemical productions used
nowadays to produce artificial fertilizers.

In previous works~\cite{CDDGGT_TCSB11,calcium_CS2BIO11} in the
context of the BioBITS project~\cite{BioBITs:web} we investigated
different phases of the AM symbiosis by using the Calculus of
Wrapped Compartments (CWC)~\cite{preQAPL2010}, a variant of the
Calculus of Looping Sequences (CLS)~\cite{BMMT06,BMMTT08}. These
calculi have been designed to represent in a formal way biological
entities such as molecules, DNA strands and cells.
Starting from an alphabet of atomic elements (\emph{atoms} for short) and from an alphabet of \emph{labels}, CWC terms are defined as multisets of atoms
and compartments. A compartment consists of a \emph{wrap} (a multiset of atoms detailing the elements of
interest on the membrane), a content (a term) and a type (a label).
The evolution of the system is driven by a set of rewrite rules modelling the reactions of interest (elements interaction and movement). In previous work,
\mc{we have introduced in CWC rules a notion of rate  which allow CWC to be equipped with a semantics given in terms of Continuous Time Markov Chains (CTMCs), thus enabling stochastic simulation built following the standard Gillespie's
approach~\cite{G77}.} Moreover, we have defined a hybrid simulation algorithm to speed up the simulation procedure while keeping the exactness of
Gillespie's method,~\cite{MECBIC10}. The simulation tool~\cite{HCWC_SIM} has been integrated in the FastFlow~\cite{fastflow:web} programming framework for
the execution of parallel simulations in multi-core platforms~\cite{ACDDTT_PDP11}.

In this paper, we continue our ongoing research on the AM symbiosis by modelling the colonization of the plant root cells by the fungal hyphae spreading in the soil.
\es{The interaction begins with a molecular dialogue between the plant and the fungus. Host roots release signalling molecules characterized as strigolactones which induce alterations in the fungal physiology and hyphal branching. In response to the plant signal AM fungi produce a symbiotic signal, the ``Myc Factor''. One of the first response observed in epidermal plant root cells is a repeated oscillation of $Ca^{2+}$ concentration preparing to the reorganization of the cell for the fungal penetration. Following this event, fungal hyphae enter and cross the epidermal cells, growing inter-and intracellularly
all along the root in order to spread fungal structures.
Once inside the inner layers of the cortical cells, specialised fungal hyphae differentiate into arbuscules where the nutrients exchange between the two organisms occur.}

\mc{To represent this kind of behaviour we need to model some spatial properties of the system like the movement of the hyphae in the soil and their penetration and articulation in the root tissue.
To this aim we exploit the fact that} the notion of compartment can also be used, \mc{in a very natural way,}  to represent spatial regions (with a fixed topology) in which the labels
play a key role in defining the spatial properties. These kind of compartments are thus considered as the topological sectors in the analysed model. In
this framework, the movement and growth of system elements are described via specific rules (involving adjacent compartments)
and the functionalities of biological components can be affected by the spatial constraints given by the sector in which they interact with other elements. \mc{A similar approach can be found in \cite{MV10} where the topological structure of the components is expressed via explicit links which require ad-hoc rules to represent movements of biological entities.}

Although our approach does not support an explicit representation of continuous space, we can exploit the topological structure of compartments,
their inter-relations, and their relative placement as a starting point to implicitly define spatial properties.  This level of abstraction is suitable to
model some situations where it is not needed to explicitly specify the space occupied by the elements of the system. Avoiding to
introduce explicit features for modeling a spatial geometry (e.g., coordinates, position, extension, motion direction and speed, rotation, collision and
overlap detection, communication range, etc.) may in some cases reduce the complexity of the models and simplify the analysis of the system and the simulation procedure.

\subsection{Summary}

The remainder of the paper is organised as follows. In Section~\ref{SecCWC} we recall the syntax and semantics of CWC. In Section~\ref{SecTI}, through a
few chosen paradigmatic examples, we start giving some hint about spatial modelling and analysis within the CWC framework. In Section~\ref{SecBio} we
exploit the approach to model some spatial interaction in the Arbuscular Mycorrhizal Symbiosis. In particular, we show how molecules (carrying the
communication signals between the plant and the fungus) diffuse in the soil, and how the fungus contacts, through its hyphae, the cells of the plant root.
Finally, in Section~\ref{SecConc} we briefly discuss related and possible future work and draw our conclusions.

\section{The Calculus of Wrapped Compartments}
\label{SecCWC}

\newcommand{\ov}[1]{\overline{#1}}
\newcommand{\srewRule}[3]{#1\srew{#3}#2}
\newcommand{\srew}[1]{\stackrel{#1}{\longmapsto}}
\newcommand{\csrew}[1]{\stackrel{#1}{\longmapsto}_{\sf c}}
\renewcommand{\wr}[3]{( #1 \into #2)^{#3}}

The Calculus of Wrapped Compartments (CWC)
(see~\cite{preQAPL2010,CDDGGT_TCSB11}) is based on a nested structure of ambients delimited by membranes with specific proprieties. Biological entities
like cells, bacteria and their interactions can be easily described in CWC.

\subsection{Syntax}
\label{CWC_formalism - syntax}

Let $\AT$ be a set of  \emph{atomic elements} (\emph{atoms} for
short), ranged over by $a$, $b$, ..., and  $\LT$ a set of \emph{compartment types} represented as \emph{labels} ranged over by $\ell,\ldots$.
A \emph{term} of CWC is a multiset $\ov{t}$ of  \emph{simple terms} where a simple term  is either an atom $a$ or a compartment $(\overline{a}\into
\overline{t'})^\ell$ consisting of a \emph{wrap}
 (represented by a multiset of atoms $\overline{a}$)\footnote{the wrap of a compartment contains the elements characterizing the membrane enclosing it. So we assume that it cannot contain other compartments but only atomic elements (for instance proteins).}, represented by a (possibly empty) and a \emph{content},represented by a term
$\overline{t'}$) and a \emph{type}, represented by the label $\ell$. An empty multiset is represented symbolically with ``$\emptyseq$''.

\begin{figure*}
\centering
\subfigure[] {
\includegraphics[height=28mm]{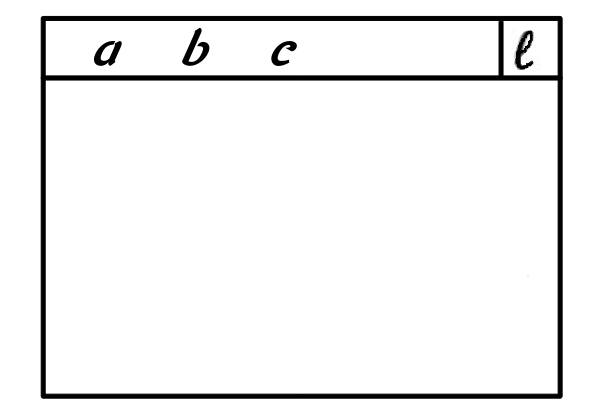}
}
\centering
\subfigure[] {
\includegraphics[height=28mm]{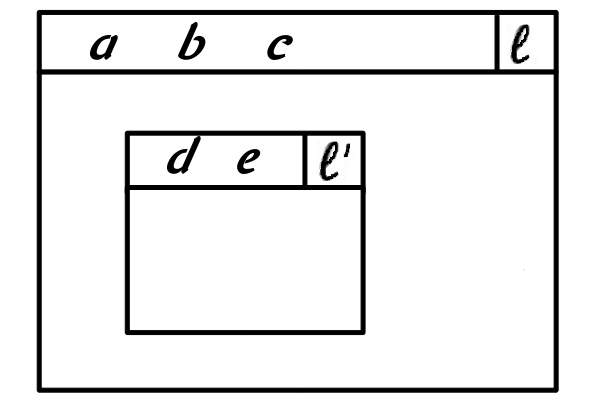}
}
\subfigure[] {
\includegraphics[height=28mm]{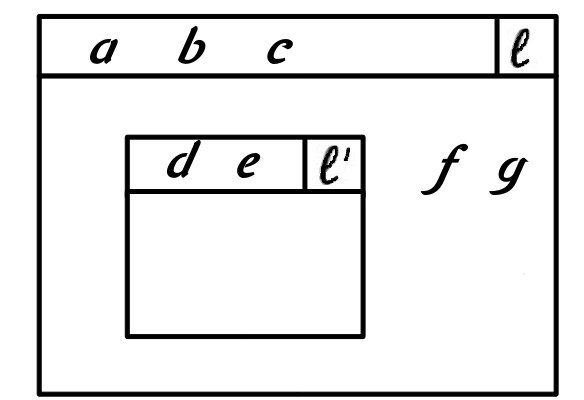}
}

\caption{\textbf{(a)} represents $(a \conc b \conc c \into \emptyseq)^\ell$; \textbf{(b)} represents $(a \conc b \conc c  \into (d \conc e \into
\emptyseq)^{\ell'})^\ell$; \textbf{(c)} represents $(a \conc b \conc c \into (d \conc e \into \emptyseq)^{\ell'} \conc f \conc g)^\ell$}
\label{fig:example CWM}
\end{figure*}
As usual, the notation $n*t$ denotes $n$ occurrences of the simple term $t$. An example of term is $2\!*\!a \conc b \conc (c \conc d \into e \conc
f)^\ell$ representing a multiset (multisets are denoted by listing the elements separated by a space) consisting of two occurrences of $a$, one occurrence
of $b$ (e.g. three molecules) and an $\ell$-type compartment $(c \conc d \into e \conc f)^\ell$ which, in turn, consists of a wrap (a membrane) with two
atoms $c$ and $d$ (e.g. two proteins) on its surface, and containing the atoms $e$ (e.g. a molecule) and $f$ (e.g. a DNA strand). See
Figure~\ref{fig:example CWM} for some other examples with a simple graphical representation.


\subsection{Rewriting Rules}
\label{CWC_formalism - rewriting}

System transformations are defined by rewriting rules written over an
extended set of terms including variables. The l.h.s. component  of a rewrite rule $\ov{\LeftPat}$ is called  \emph{pattern} and  the r.h.s. component $ \ov{\RightPat}$ of a rewrite rule is called \emph{open term}.  Patterns and open terms are multiset of \emph{simple patterns} $p$ and \emph{simple open terms} $o$ defined by the following syntax:
$$
\begin{array}{lcl}
   \LeftPat  & \;\qqop{::=}\; & a \agr (\overline{a} \conc x \into \overline{\LeftPat}\conc X)^\ell \\
   \RightPat & \;\qqop{::=}\; & a \agr (\ov{a}  \conc x \into \ov{o})^\ell   \agr X
\end{array}
$$

where $\overline{a}$ is a multiset of atoms, $x$ and $X$ are
variables that can be instantiated, respectively by a multiset of atoms (on a wrap) or by a term (in the  compartment contents). The label $\ell$ is called the \emph{type of the pattern}. Patterns are intended to match, via substitution of variables, with simple terms and compartments occurring as subterms of the term representing the whole system. Note that we force \emph{exactly} one variable to occur in each
compartment content and wrap.
 This prevents ambiguities in the instantiations needed to match a given compartment.
 \footnote{
 The linearity condition, in biological terms, corresponds to the exclusion of transformations depending on the presence of two (or more)
 identical (and generic) components in different compartments (see also~\cite{OP11}).}

A \emph{rewrite rule} is a triple $(\ell, \ov{\LeftPat},\ov{\RightPat} $, denoted by $\ell:  \overline{\LeftPat} \srew{}    \ov{\RightPat})$, where $\ov{\LeftPat}$ and $\ov{\RightPat}$ are such that 
the variables occurring in $\ov{\RightPat}$ are a subset of the variables occurring in $\ov{\LeftPat}$.
 %
The application of a rule $\ell:  \ov{\LeftPat} \red \ov{\RightPat}$ to a term $\ov{t}$ is performed in the following way: 1) Find in $\ov{t}$ (if it
exists) a compartment of type $\ell$ with content $\ov{u}$ and a substitution $\sigma$ of variables by ground terms (i.e. terms without occurrences of variables) such that $\ov{u} = \sigma( \ov{\LeftPat} \conc
X)$, where $X$ is a variable not occurring in $\ov{p},~ \ov{o}$; and 2) Replace in $\ov{t}$ the
subterm $u$ with $\sigma(\ov{\RightPat}\conc X)$. We write  $\ov{t} \red \ov{t'}$ if $\ov{t'}$ is obtained  by applying a rewrite rule to  $\ov{t}$.

For instance, the rewrite rule
  $\ell:~a \; b \red c$
means that in all compartments of type $\ell$ containing an occurrence of $a \conc b$, it can be replaced by $c$. While the rewrite rule  $\ell: (x \into a \conc b \conc X)^{\ell_1} \red ( x \into c \conc X)^{\ell_2}$ means that, if contained in a compartment of type $\ell$, all compartments of type $\ell_1$ and containing an $a \conc b$ in their content can change their type to $\ell_2$ and an occurrence of $a \conc b$  can be replaced by $c$.

 \noindent \emph{Remark.} For  uniformity we assume that the term representing the whole
system is always a single compartment labelled $\top$ with an empty wrap, i.e., all systems are represented by a term of the shape
$\wr{\emptyseq}{\ov{t}}{\top}$, which we will also write as $\ov{t}$ for simplicity.

\subsection{Stochastic Simulation}\label{SECT:STO_SEM}

A stochastic simulation model for biological systems can be defined
along the lines of the one
presented by Gillespie in \cite{G77}, which is, \emph{de facto}, the standard way to model quantitative aspects of biological systems. The basic idea of
Gillespie's algorithm is that a rate function is associated with each considered chemical reaction which is used as the parameter of an exponential distribution modelling the probability that the reaction takes place. In the standard approach this reaction rate is obtained by multiplying the kinetic
constant of the reaction by the number of possible combinations of reactants that may occur in the region in which the reaction takes place, thus modelling the law of mass action. In \cite{preQAPL2010}, the reaction rate is defined in a more general way by associating to each reduction rule a function which can also define rates based on different principles as, for instance, the Michaelis-Menten nonlinear kinetics.

In the examples of this paper it is reasonable to assume a linear dependence of the reaction rate on the number of possible combinations of interacting components, so we will follow the standard approach in defining reaction rates. Each reduction rule is then enriched by the kinetic constant $k$ of the reaction that it represents (notation $\ell:\LeftPat \srewrites{k} \RightPat $).
For instance in evaluating the application rate of the stochastic rewrite rule $R=
\ell: a \conc b \conc X \srewrites{k} c \conc X$ to the term $\ov{t}=a\conc a \conc b \conc b$ in a compartment of type $\ell$ we must consider the number of the possible combinations of reactants of
the form $a\conc b$ in $\ov{t}$. Since each occurrence of $a$ can react with each occurrence of $b$, this number is 4. So the application rate of $R$ is
$k\cdot 4$.

%

\subsection{The CWC simulator}
\label{sec:simulator}

The CWC simulator~\cite{HCWC_SIM} is a tool under development at the Computer science Department of Turin University, based on Gillespie's direct method
algorithm~\cite{G77}. It treats CWC models with different rating semantics (law of mass action, Michaelis-Menten kinetics, Hill equation) and
it can run independent stochastic simulations over CWC models, featuring deep parallel optimizations for multi-core platforms on the top of
FastFlow~\cite{fastflow:web}. It also performs online analysis by a modular statistical framework.


\section{Topological Interpretations within CWC}
\label{SecTI}
In this section we provide some hint about spatial modelling and analysis within the CWC framework by means of some paradigmatic examples. For simplicity,
we will consider the compartments in our system to be \emph{well-stirred}. Thus, also the compartments representing the spatial sectors of our topology
will consist of well-mixed components. Note that the topological analysis described in the following examples could be expressed also with other calculi
with compartmentalisation such as, for example, BioAmbients~\cite{RPSCS04}, Brane Calculi~\cite{Car04} and Beta-Binders~\cite{DPPQ06}.

\subsection{Cell Growth and Proliferation: A CWC Grid}\label{SECT_GRID}

Eukaryotic cells reproduce themselves by cell cycle which can be divided in two brief periods: the interphase during which the cell grows, accumulating nutrients and duplicating its DNA and the mitosis ($M$) phase, during which the cell splits itself into two distinct cells, often called ``daughter cells''.
Interphase proceeds in three stages, $G1$, $S$, and $G2$. $G1$ (Gap1) is also called the growth phase, and is marked by synthesis of various enzymes that are required in $S$ phase, mainly those needed for DNA replication. In $S$ phase (Synthesis) DNA replication occurs. The cell then enters the $G2$ (Gap2) phase, which lasts until the cell enters mitosis. Again, significant biosynthesis occurs during this phase, mainly involving the production of microtubules, which are required during the process of mitosis.


Spatial representation of the growth of cell populations could be useful in many situations (in the study of bacteria colonies, in developmental biology, and in the analysis of cancer cells proliferation).

Intuitively, we may express cell proliferation by describing a finite space through a CWC grid such that on each  \emph{grid cell} we can insert a cell of the growing population under analysis. An $k\times n$ grid could be modelled in CWC with $k*n$ compartment labels modelling the \emph{grid cells}. A cell may grow in one of the adjacent \emph{grid cells} (with a given scheme of adjacency modelled by the set of rewrite rules).

Intuitively, using the atomic element $e$ to denote an empty \emph{grid cell}, an empty grid could be modelled by the following CWC term:

$$
\begin{array}{ll}
\texttt{Prol\_Grid}= & (  \into e )^{1,1} \conc \ldots \conc (  \into e )^{1,n} \conc\\
& (  \into e )^{2,1} \conc \ldots \conc (  \into e )^{2,n} \conc\\
& \ldots \\
& (  \into e )^{m,1} \conc \ldots \conc (  \into e )^{m,n} \conc\\
\end{array}
$$

If we represent a cell of our population as a compartment with membrane $m$ and with a single atom in its content representing the phase of the cell cycle ($M$, $G1$, $S$, $G2$), a \emph{grid cell} $i,j$ may contain the cell $(m \into M)^{cell}$ that could grow, with kinetic $k_{mit}$, towards the empty \emph{grid cell} $i,j+1$ with the rewrite rule
\footnote{note that under our assumptions the variables 
will always match only with the empty multiset}:

$$
\begin{array}{lcrl}
&\TOP & : & ( x \into  ( m \conc y \into M \conc Y)^{cell} \conc X )^{i,j} ( z \into  e \conc Z )^{i,j+1} \srewrites{k_{mit}} ( x \into  ( m \conc y \into G1 \conc Y)^{cell} \conc X  )^{i,j} ( z \into  ( m \into G1)^{cell} \conc Z )^{i,j+1} \\
\end{array}
$$

\noindent  while cells may change their internal state (cell cycle phase) with the rules:

$$
\begin{array}{lcrl}
&i,j & : & ( m \conc x \into G1 \conc X)^{cell}  \srewrites{k_{phase1}}  ( m \conc x \into S \conc X)^{cell}\\
&i,j & : & ( m \conc x \into S \conc X)^{cell}  \srewrites{k_{phase2}}  ( m \conc x \into G2 \conc X)^{cell}\\
&i,j & : & ( m \conc x \into G2 \conc X)^{cell}  \srewrites{k_{phase3}}  ( m \conc x \into M \conc X)^{cell}
\end{array}
$$

Note that each element of the grid can be occupied only by one cell, so this model allows only a limited growth of the cells, corresponding to the dimension of the grid. However this could be enough to investigate the properties of the cell growth.

\subsection{Quorum Sensing and Molecular Diffusion}\label{SECT_QS}

Molecular diffusion (or simply diffusion) arises form the motion of particles as a function of temperature, viscosity of the medium and the size (or mass) of the particles. Usually, diffusion explains the flux of molecules from a region of higher concentration to one of lower concentration, but it may also occur when there is no concentration gradient. The result of the diffusion process is a gradual mixing of the involved particles. In the case of uniform diffusion (given a uniform temperature and absent external forces acting on the particles) the process will eventually result in a complete mixing.
Equilibrium is reached when the concentrations of the diffusing molecules between two compartments becomes equal.

\begin{figure}
\begin{center}
\begin{minipage}{0.98\textwidth}
\begin{center}
\includegraphics[width=80mm]{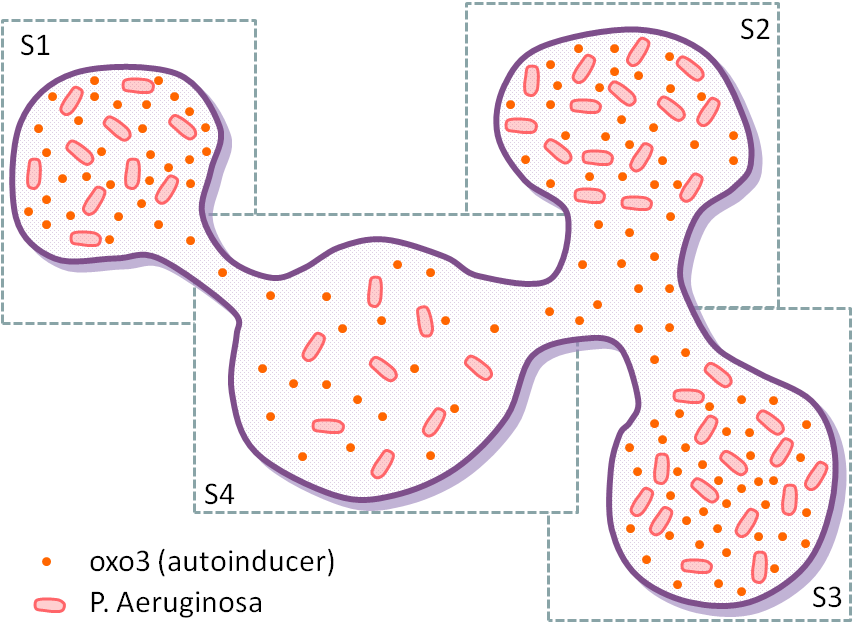}
\end{center}
\caption{\label{FIG_QS}Quorum Sensing in a distributed topology.}
\end{minipage}
\end{center}
\end{figure}
Diffusion is of fundamental importance in many disciplines of physics, chemistry, and biology.
In this subsection we will consider the diffusion of the auto-inducer communication signals in a quorum sensing process where the bacteria initiating the process are distributed in different sectors of an environment (see Figure~\ref{FIG_QS}). We consider the sectors of the environment to be connected through channels with different capacities and different speeds regulating the movement of the auto-inducer molecules between the different sectors.

\begin{figure}
\center
\includegraphics[width=.49\textwidth]{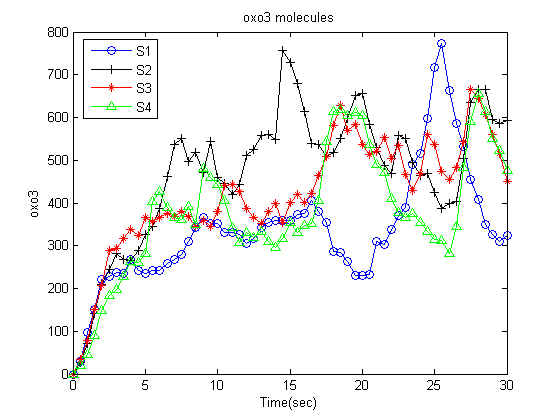}
\includegraphics[width=.49\textwidth]{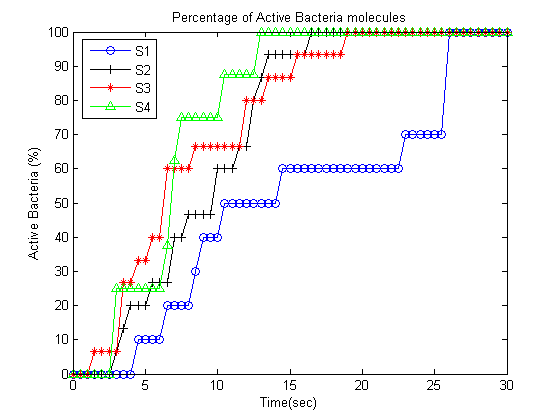}
\caption{\label{res_qs} Simulation results of oxo3 molecules (left figure) and active bacteria percentage with respect to the total number of bacteria (right figure) in the different sectors.}
\end{figure}
The CWC model describing the system can be found in Appendix A. We modelled the four sectors as four different compartments with special labels $Si$ allowing the flow of the oxo3 auto-inducer molecules with different speed. Bacteria (modelled as compartments with label $m$) were marked as ``active'' when they reached a sufficient high level of the oxo3 auto-inducer. The results of a simulation running for $30$ seconds is shown in Figure~\ref{res_qs}.

Although CWC does not have a direct mechanism to formulate the spatial position of its terms and a metric distance between them, it has the potential to describe topologically different spatial concepts. For example, the compartments are able, through the containment specification, to express discrete distance among the entities that are inside and those on the outside. Moreover, through the rule kinetics we are able to model space anisotropy (lack of uniformity in all orientations), which are useful to formulate complex concepts of reachability. The model discussed in this section is a clear example of these issues.

\section{Spatial Interactions in the Arbuscular Mycorrhizal Symbiosis}
\label{SecBio}
Arbuscular mycorrhizae, formed between $80\%$ of land plants and arbuscular mycorrhizal (AM) fungi belonging to the Glomeromycota phylum, are the most common and widespread symbiosis on our planet \cite{harrison2005signaling}. AM fungi are obligate symbionts 
which supply the host with essential nutrients such as phosphate, nitrate and other minerals from the soil. In return, AM fungi receive carbohydrates derived from photosynthesis in the host. AM symbiosis also confers resistance to the plant against pathogens and environmental stresses. Despite the central importance of AM symbiosis in both agriculture and natural ecosystems, the mechanisms for the formation of a functional symbiosis between plants and AM fungi are largely unknown.

\subsection{The Biological Interactions under Investigation}

The interaction begins with a molecular dialogue between the plant and the fungus \cite{harrison2005signaling}. Host roots release signalling molecules characterized as strigolactones \cite{akiyama2005plant} which are sesquiterpenes derived from the carotenoid pathway \cite{matusova2005strigolactone,lopez2008tomato} 
Within just a few hours, strigolactones at subnanomolar concentrations induce alterations in fungal physiology and mitochondrial activity and extensive hyphal branching \cite{besserer2006strigolactones}.
Root exudates from plants grown under phosphate-limited conditions are more active than those from plants with sufficient phosphate nutrition, suggesting that the production of these active exudates in roots is regulated by phosphate availability \cite{nagahashi2000partial,yoneyama2001production}.
Strigolactones also appear to act as chemo-attractants: fungal hyphae of \textit{Glomus mosseae} exhibited chemotropic growth towards roots at a distance of at least 910 $\mu$m in response to host-derived signals, possibly strigolactones \cite{sbrana2005chemotropism}.
Some solid evidence has been presented for AM fungal production of a long-hypothesized symbiotic signal, the ``Myc Factor'', in response to the plant symbiotic signal. Fungal hyphae of the genus \textit{Gigaspora} growing in the vicinity of host roots, but separated from the roots by a membrane, release a diffusible substance that induces the expression of a symbiosis-specific gene 
in \textit{Medicago truncatula} roots \cite{kosuta2003diffusible}. This expression was correlated both spatially and temporally with the appearance of hyphal branching, and was not observed when hyphal branching was absent.
The external signal released by these fungi is perceived by a receptor on the plant plasma membrane and is transduced into the cell with the activation of a symbiotic signalling pathway that lead to the colonization process.
One of the first response observed in epidermal cells of the root is a repeated oscillation of $Ca^{2+}$ concentration in the nucleus and perinuclear cytoplasm through the alternate activity of $Ca^{2+}$ channels and transporters \cite{oldroyd2008coordinating,kosuta2008differential,chabaud2011arbuscular}.
Kosuta and collegues \cite{kosuta2008differential} assessed mycorrhizal induced calcium changes in \textit{Medicago truncatula} plants transformed with the calcium reporter cameleon. Calcium oscillations were observed in a number of cells in close proximity to plant/fungal contact points. Very recently Chabaud and colleagues \cite{chabaud2011arbuscular} using root organ cultures of both \textit{Medicago truncatula} and \textit{Daucus carota} 
observed $Ca^{2+}$ spiking in AM-responsive zone of the root treated with AM spore exudate.

Once reached the root surface the AM fungus differentiates a hyphopodium via
which it enters the root. 
A reorganization of the plant cell occurs before the fungal penetration: a thick cytoplasmic bridge (called pre-penetration apparatus -PPA-) 
is formed and it constitutes a trans-cellular tunnel in which
the fungal hypha will grow \cite{genre2005arbuscular}. Following this event, a hyphal peg is produced which enters and crosses the epidermal cell, avoiding a direct contact between fungal
wall and host cytoplasm thanks to a surrounding membrane of host origin \cite{genre2007check}.
Than the fungus overcomes the epidermal layer and it grows inter-and intracellularly
all along the root in order to spread fungal structures.
Once inside the inner layers of the cortical cells the differentiation of specialized, highly branched intracellular hyphae called arbuscules occur. Arbuscules are considered the major site for nutrients exchange between the two organisms 
and they form by repeated dichotomous hyphal branching and grow
until they fill the cortical cell. They are ephemeral structures \cite{toth1984dynamics,javot2007medicago}: at some point after maturity, the arbuscules collapse, degenerate, and die and the plant cell regains its previous organization \cite{bonfante1984anatomy}. The mechanisms underlying arbuscule turnover are unknown
but do not involve plant cell death. In arbuscule-containing cells the two partners come in intimate physical contact and adapt their metabolisms to allow reciprocal benefits.

\subsection{The CWC Model}
We model the Arbuscular Mycorrhyzal symbiosis in a 2D space. The soil environment is partitioned into different concentric layers to account for the distance between the plant root cells and the fungal hyphae where the inner layer contains the plant root cells. The plant root tissue is also modelled as concentric layers each one composed by sectors.

\begin{figure}
\center
\includegraphics[width=.5\textwidth]{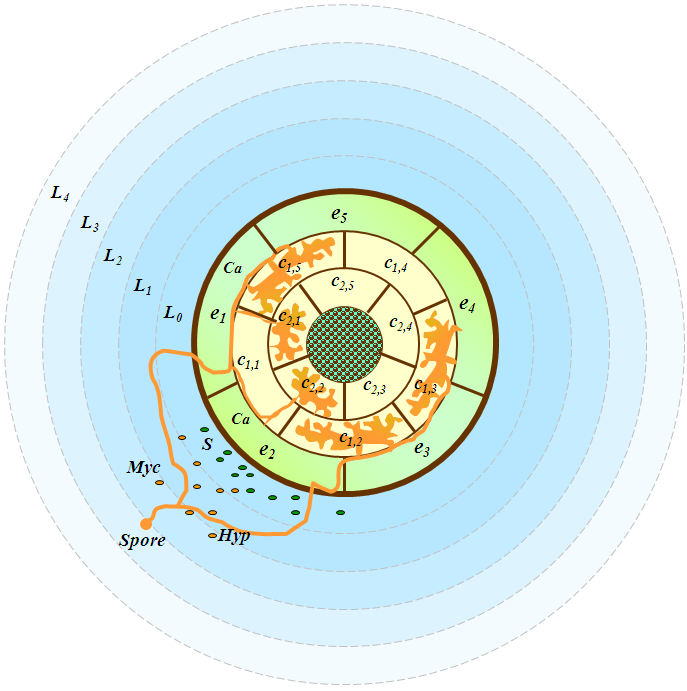}
\caption{\label{sym_schema}2D spatial model of the Arbuscular Mycorrhyzal symbiosis.}
\end{figure}

For the sake of simplicity we modelled the soil into $5$ different concentric layers (modelled as different compartments labelled by $L_0$-$L_{4}$ identifying $L_4$ with the top level $\TOP$). \es{Although the root has a complex layered structure we simplified it} into $3$ concentric layers (one epidermal layer and $2$ cortical layers) each one composed by $5$ sectors mapped as different compartments labelled by $e_i$ ($i=1 \ldots 5$) for epidermal cells and $c_{jk}$ ($j=1, 2$ and $k=1 \ldots 5$) for cortical cells. Exploiting the flexibility of CWC, compartment labels are here used to characterize both their spatial position and their biological properties. All the compartments and the main atoms of the model are depicted in Figure~\ref{sym_schema}.

The stigolactones (atoms $S$) derived from the carotenoid pathway (atom $C_P$) are released by epidermal cells and can diffuse through the soil degrading their activity when farthest from the plant. This situation was modelled by the following rules:

\begin{equation}
\begin{array}{lcrlr}
(R_{CP_i}) & L_0 & : &  ( x \into C_P \conc X)^{e_i} \srewrites{K_{CP}} S ( x \into C_P \conc X)^{e_i} & i=1 \ldots 5\\
(R_{S_i})  & L_{i+1} & : &  ( x \into S \conc X)^{L_i} \srewrites{K_{S}} S ( x \into \conc X)^{L_i} &  i=0 \ldots 3\\
(R_{S_{4}})  & \TOP & : &  S \srewrites{K_{S}}  \emptyseq &
\end{array}
\end{equation}

Since phosphate (atoms $P$) conditions regulate strigolactones biosynthesis inhibiting the carotenoid pathway (atom $IC_P$), the following rules were considered :

\begin{equation}
\begin{array}{lcrlr}
(R_{ICP_i}) & L_0 & : & P ( x \into C_P \conc X)^{e_i} \srewrites{K_{ICP}} P ( x \into IC_P \conc X)^{e_i} & i=1 \ldots 5
\end{array}
\end{equation}

The fungal structure is described by atoms
where each atom represents a particular type of fungal component. The atom $Spore$ represents the fungal spore while atoms $Hyp$ represent fungal hyphae.
AM fungal hyphae branching towards the plant root is regulated by the strigolacones activity. If we locate the fungal spore at a fixed distance from the plant root (let us say at the layer $L_k$ where $0 < k < 4$) then the spore hyphal branching is generated stochastically in the relative inferior and superior layers. In our experiments the fungal spore was placed at the soil layer $L_2$. The hyphal branching is radial with respect to the spore position causing a branching away from the spore and promoted at the proximity of the root by the strigolactones. This branching process was represented by the following rules:

\begin{equation}
\begin{array}{lcrlr}
(R_{F_1}) & L_2 & : & Spore ( x \into  \conc X)^{L_{1}} \srewrites{K_{F}}  Spore ( x \into Hyp \conc X)^{L_{1}}  & \\
(R_{F_2}) & L_{3} & : & ( x \into Spore \conc X)^{L_{2}} \srewrites{K_{F}}  Hyp ( x \into Spore \conc X)^{L_{2}}  & \\
(R_{H_j}) & \TOP & : & ( x \into Hyp \conc X)^{L_{3}} \srewrites{K_{H}}  Hyp ( x \into Hyp \conc X)^{L_{3}}  & \\
(R_{H_1}) & L_{1} & : & S  \conc Hyp ( x \into \conc X)^{L_{0}} \srewrites{K_{H_1}}  S  \conc Hyp ( x \into Hyp \conc X)^{L_{0}}  &
\end{array}
\end{equation}

The AM fungal production of Myc factor (atom $Myc$) at the proximity of the root cells inducing the calcium spiking phenomenon (atom $Ca$) at a responsive state (atom $Resp$) which prepares the epidermal cells to include the fungal hyphae was modelled by the following rules:

\begin{equation}
\begin{array}{lcrlr}
(R_{Myc}) & L_0 & : & Hyp \srewrites{K_{M}} Myc \conc Hyp & \\
(R_{MycCa_i}) & L_0 & : & Myc ( Resp \conc x \into \conc X)^{e_i} \srewrites{K_{MC}}  (Resp \conc x \into \conc Ca \conc X)^{e_i} & i=1 \ldots 5\\
(R_{Ca_i}) & L_0 & : & Hyp ( Resp \conc x \into Ca \conc X)^{e_i} \srewrites{K_{Ca}} Hyp ( NResp \conc x \into Hyp \conc X)^{e_i} & i=1 \ldots 5
\end{array}
\end{equation}

where the epidermal cells change their state from responsive to not responsive (atom $NResp$) since they can be penetrated by only one hypha.

Once the fungal hyphae penetrate the epidermal cells they are allowed to branch in the two directions of the underneath cortical layer and from this  cortical layer to the inner one. The cortical cells change their state from responsive (atom $Resp$) to transitive (atom $Trans$) and then to mycorrhized (atom $Arb$) since an hypha can grow in the cortical intercellular spaces even if already mycorrhized but only one arbuscule can be generated. The hyphal branching from the epidermal layer to the cortical layers is modelled through the following rules:

\begin{equation}
\begin{array}{lcrlr}
(R_{CB_{1i}}) & L_0 & : & ( NResp \conc x \into Hyp \conc X)^{e_i} ( Resp \conc y \into \conc Y)^{c_{1j}} \srewrites{K_{CB}}  ( NResp \conc x \into Hyp \conc X)^{e_i} ( Trans \conc y \into  Hyp \conc Y)^{c_{1j}} & \\
& & & i=1 \ldots 5, j=i (mod 5), i+1 (mod 5); & \\
(R_{CB1_{2i}}) & L_0 & : & ( Trans \conc x \into Hyp \conc X)^{c_{1i}} ( Resp \conc y \into  \conc Y)^{c_{2j}} \srewrites{K_{CB}}  ( Trans \conc x \into Hyp \conc X)^{c_{1i}} ( Trans \conc x \into  Hyp \conc Y)^{c_{2j}} & \\
 & & & i=1 \ldots 5, j=i (mod 5), i+1 (mod 5); & \\
(R_{CB2_{2i}}) & L_0 & : & ( Arb \conc x \into Hyp \conc X)^{c_{1i}} ( Resp \conc y \into  \conc Y)^{c_{2j}} \srewrites{K_{CB}}  ( Arb \conc x \into Hyp \conc X)^{c_{1i}} ( Trans \conc y \into  Hyp \conc Y)^{c_{2j}} & \\
 & & & i=1 \ldots 5, j=i (mod 5), i+1 (mod 5); & \\
\end{array}
\end{equation}

Finally, the modelling of each fungal hyphae penetrated into a cortical cell to create an arbuscule (atom $A$) and start the symbiosis is performed by the following rules:

\begin{equation}
\begin{array}{lcrlr}
(R_{A_{ij}}) & L_0 & : & ( Trans \conc x \into Hyp \conc X)^{c_{ij}} \srewrites{K_{A}}   ( Arb \conc x \into Hyp \conc X)^{c_{ij}} & i=1, 2 ; j=1 \ldots 5\\
\end{array}
\end{equation}

\subsection{Simulation Results}

The initial term describing our system is given by $T$:
\begin{equation}
\begin{array}{lll}
T =& ( \into  \conc ( \into  \conc ( \into Spore \conc ( \into  \conc ( \into  \conc T')^{L_{0}})^{L_{1}})^{L_{2}})^{L_{3}})^{\TOP}\\
&\text{where} \\
&T'= V \conc U \\
& V = V_1 \conc  V_2 \conc  V_3 \conc  V_4 \conc  V_5\\
& V_i = ( Resp \into  \conc C_P)^{e_{i}}\\
& U = U_{11} \conc  U_{12} \conc  U_{13} \conc  U_{14} \conc  U_{15} \conc  U_{21} \conc   U_{22} \conc  U_{23} \conc  U_{24} \conc  U_{25}\\
& U_{jk} = ( Resp \into  \conc )^{c_{jk}}\\
\end{array}
\end{equation}
\begin{figure}
\center
\includegraphics[width=.49\textwidth]{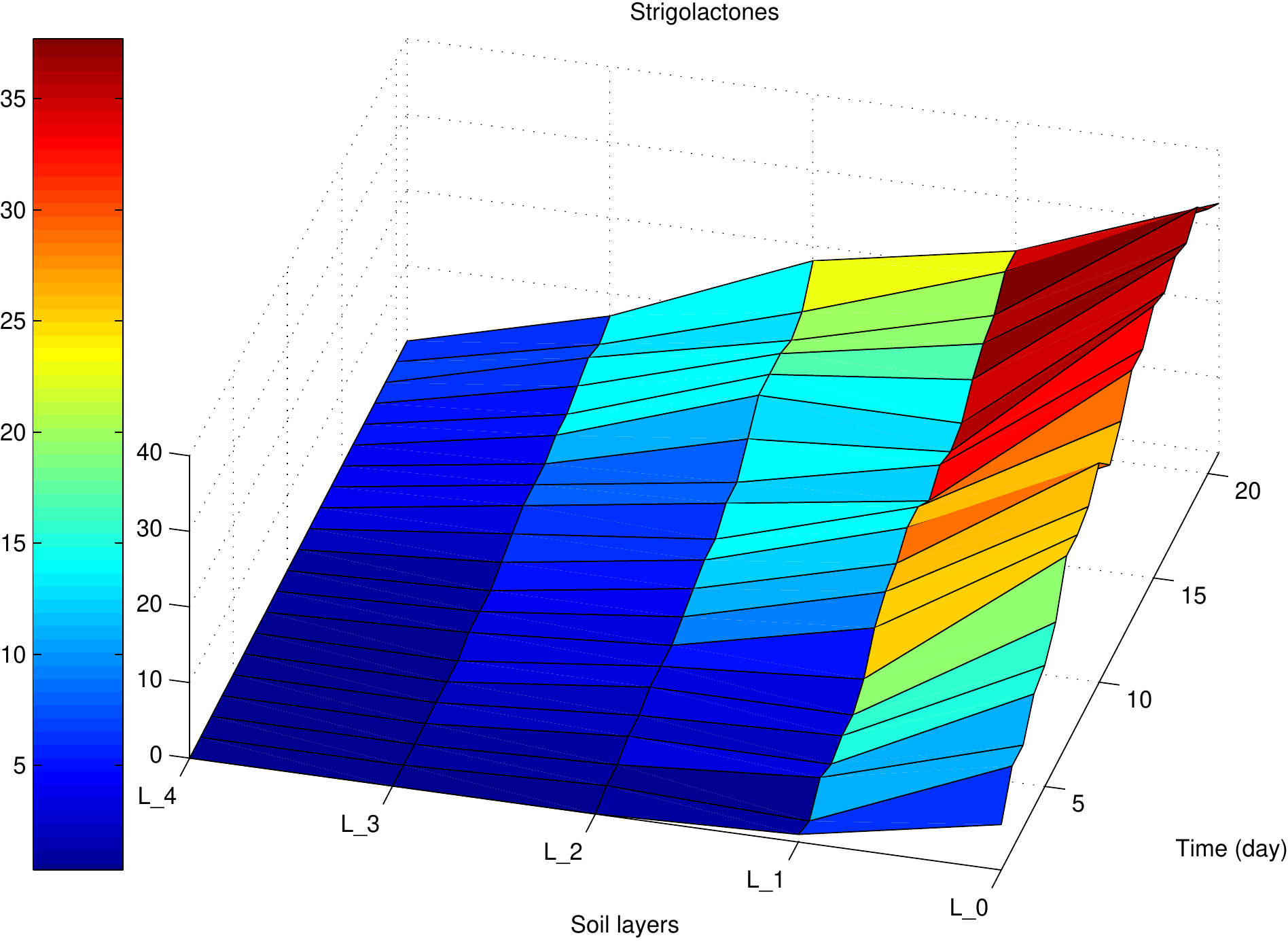}
\includegraphics[width=.49\textwidth]{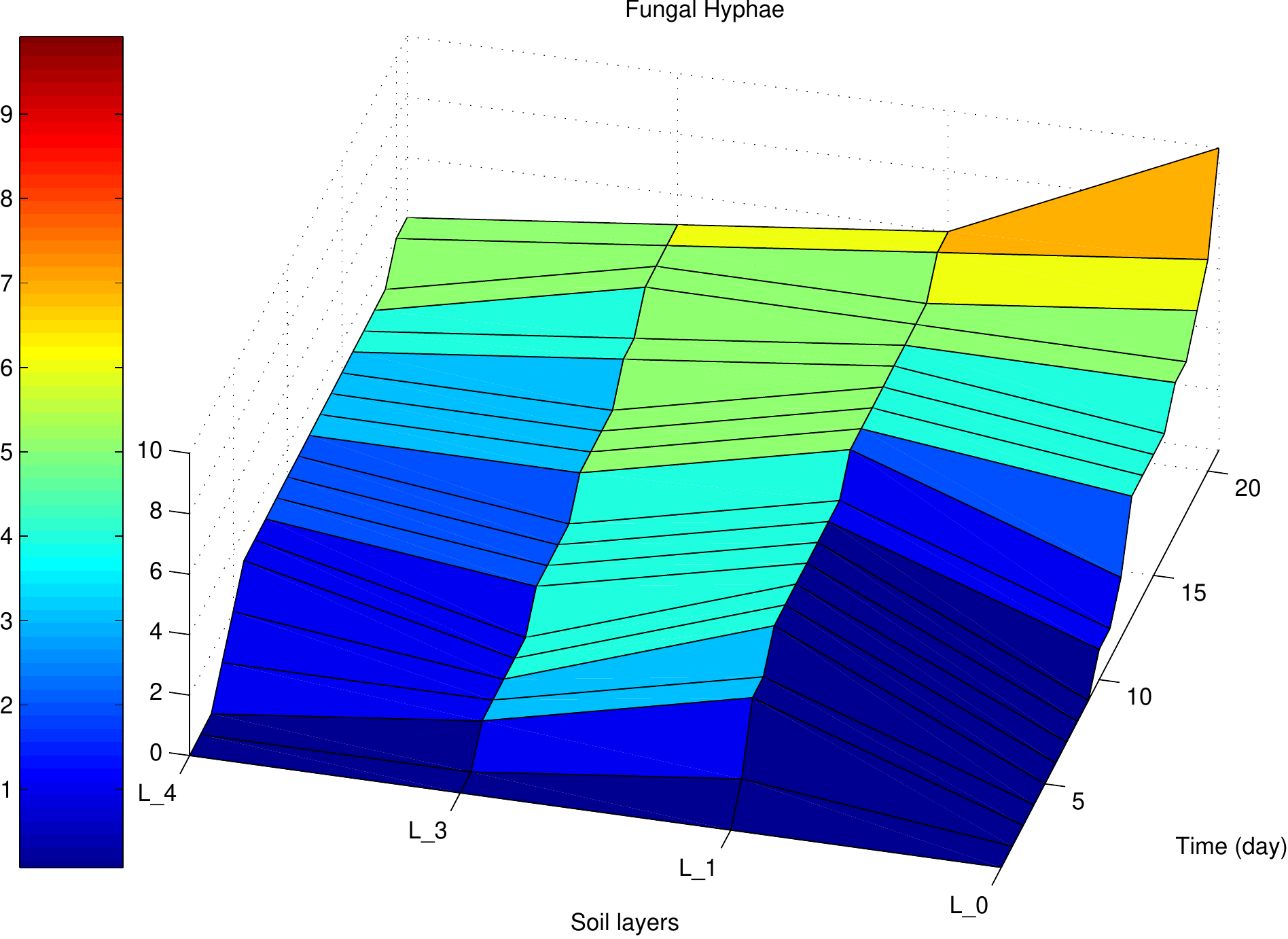}
\caption{\label{res_noP}Simulation results of strigolactones diffusion (left figure) and fungal hyphal branching (right figure) along the soil layers during time.}
\end{figure}

The model has been used to virtually simulate $21$ days of the fungus--plant interactions. \es{The reaction rates have been chosen to resemble the experimental observations of ``in vivo'' fungal germination, hyphal branching and arbuscules creation under similar environmental conditions of the simulations. The rates were set as following: $K_{CP}= 0.8$, $K_{S}=0.1$, $K_{ICP}= 0.01$, $K_{F}=0.2$, $K_{H}=0.05$, $K_{H_1}
0.01$, $K_{M} = 0.2$, $K_{MC} = 0.2$, $K_{Ca} = 1.0$, $K_{CB} = 0.9$, $K_{A} = 1.0$.}
In Figure \ref{res_noP} we show a surface plot representing an exemplification of the CWC simulation about strigolactones diffusion (on the left) and fungal hyphal branching (on the right) along the soil layers during time.

\begin{figure}
\center
\includegraphics[width=.49\textwidth]{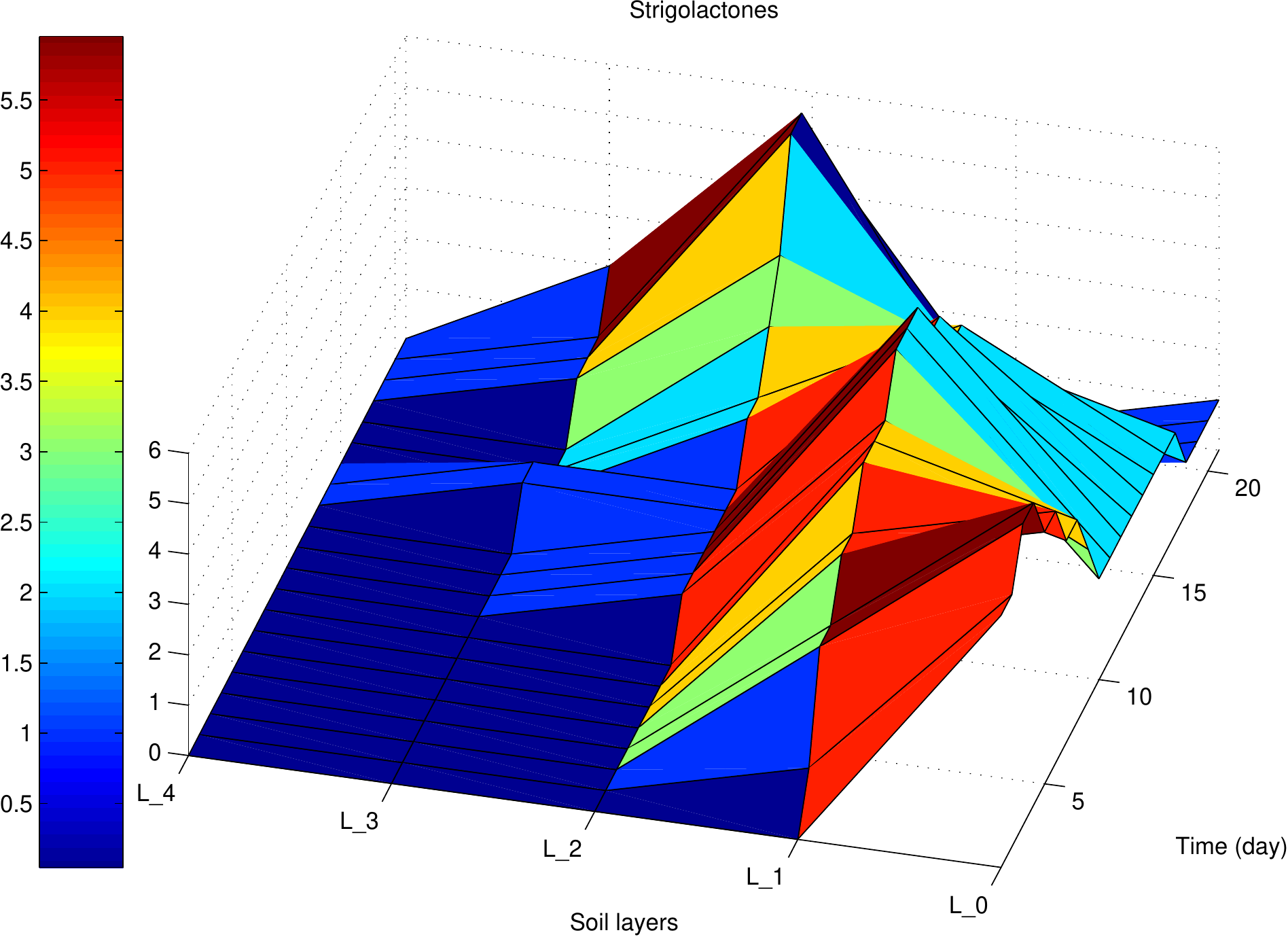}
\includegraphics[width=.49\textwidth]{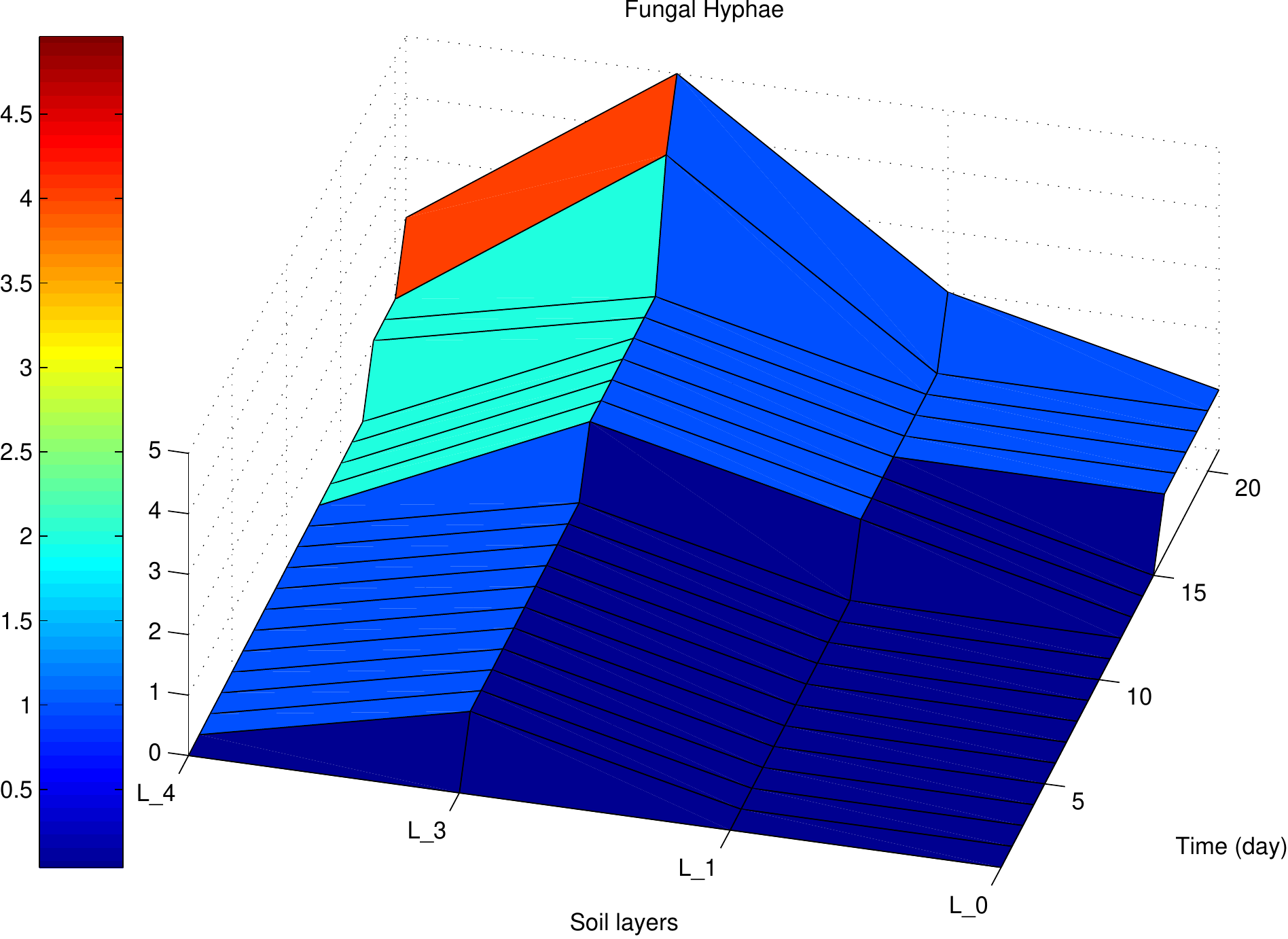}
\caption{\label{res_highP}Simulation results of strigolactones diffusion (left figure) and fungal hyphal branching (right figure) along the soil layers at high phosphate concentrations.}
\end{figure}

We compared our simulation results adding the phosphate ($100$ $P$ atoms) to the soil layers yielding to the results given by Figure \ref{res_highP}. In this case we have a lower strigolactones production by the root cells and an hyphal branching not promoted in the direction of the root.

We performed $60$ simulations in the three conditions of low phosphate ($10$ $P$ atoms), medium phosphate ($50$ $P$ atoms) and high phosphate ($100$ $P$ atoms) conditions leading to a $60\%$ of arbuscules in the first case, $30\%$ in the second case, and $10\%$ in the latter case. The first arbuscules occur around the $8^{th}-9^{th}$ day of simulation at low phosphate condition while they are formed from the $9^{th}$ to the $15^{th}$ day at high phosphate conditions.
These results reflect the in vivo experiments on several plants inoculated with different AM fungi \cite{mosse1973advances,asimi1980influence,hepper1983effect}.

Our long term goal is to design a virtual spatial model that would enable us to test ``in silico'' various
hypotheses about the interaction between the physiological processes that drive the arbuscular mycorrizal symbiosis and the signalling processes that take place in this symbiosis.

\section{Conclusions, Related and Further Work}
\label{SecConc}
In the last few years, computer scientists have developed several
languages and tools to build and analyse models of biological
systems, often relating different behaviour that operates on
different levels of abstraction and on different time and spatial
scales~\cite{PRSS01,P02,RS02,DL04,Car04,DPPQ06,KMT08}.

This work constitutes a first step towards providing CWC with the features required by the emerging field of \emph{spatial systems biology}, which aims at integrating the analysis of biological systems with spatial properties.

For the well-mixed chemical systems (even divided into nested
compartments) often found in cellular biology, interaction and
distribution analysis are sufficient to study the system's
behaviour. However, there are many other situations, like in cell
growth and developmental biology where dynamic spatial
arrangements of cells determines fundamental functionalities,
where a spatial analysis becomes essential. Thus, a realistic
modelling of many cellular phenomena requires
space to be taken into account~\cite{K06}.

There are several cases in which the spatial properties of a system could be analysed by just focusing on a finite (and quite limited) number of sectors. If we focus ourselves in this kind of situations, CWC appears to be expressive enough for quite effective descriptions. Indeed, the considerations presented in this paper about the topological organisation of the paradigmatic examples shown in Section~\ref{SecTI}, also hold for other calculi which are able to express compartmentalisation (see, e.g., BioAmbients~\cite{RPSCS04}, Brane Calculi~\cite{Car04}, Beta-Binders~\cite{DPPQ06}, etc.).

Von Neumann' Cellular Automata~\cite{vN66} were developed as a computational tool inspired by biological behaviour and later used to model biological systems. Cellular Automata are defined by a finite grid of cells, with the state of each cell, evolving in discrete time, affected locally by just the state of the nearby cells. We can mimic the framework of Cellular Automata within CWC by defining a grid in a way similar to what we have done in Section~\ref{SECT_GRID}.

There are other situations, however, where a geometrical distance becomes necessary beyond a purely topological organization (for example in a more detailed analysis in the field of developmental biology).

As a consequence, the need of an extended spatial analysis in systems biology has brought to the extension with (even continuous) spatial features of many formalisms developed for the analysis of biological systems: see, e.g.,  geometric process algebra~\cite{CG10}, spatial P-systems~\cite{BMMPT11}, spatial CLS~\cite{BMMP09} and Bioshape~\cite{BCCMT10}.

Theoretically, building a more complex structure over the approach presented in this paper, we could define a geometric space, with coordinates and distance metrics. We might define, e.g., on the lines of the approach presented in~\cite{MV10}, a surface language modelling a 2D grid or a 3D space over the CWC model. While adding too many features to the model (e.g., coordinates, position, extension, motion direction and speed, rotation, collision and overlap detection, communication range, etc.) could rise heavily the complexity of this kind of approach, a detailed study on a subset of these features could still be carried out on several systems. We plan to investigate this line of development in near future work.

\bibliographystyle{eptcs}
\bibliography{fmb,am}

\begin{appendix}
\section*{Appendix A: Quorum Sensing and Molecular Diffusion Model}
\label{appDiff}
\emph{Pseudomonas aeruginosa} uses quorum sensing to keep low the expression of virulence factors until the colony has reached a certain density, when an autoinduced production of virulence factors is started. The bacterium membrane, denoted as $m$, contains only a DNA strand. The DNA is modelled as a sequence of genes lasO lasR lasI (atom $LasORI$), where lasO represents the target to which a complex autoinducer/R-protein binds to promote transcription. The following rewrite rules model the system behaviour (the kinetics are taken from~\cite{BMMTT08}):

\begin{equation}
\begin{array}{lcrlclcrl}
(R_{1}) & m &:& LasORI  \srewrites{20} LasORI  \conc  LasR  &{}& (R_{2}) & m &:& LasORI   \srewrites{5} LasORI \conc LasI   \\
(R_{3}) & m &:& LasI   \srewrites{8} LasI \conc  oxo3   &{}& (R_{4}) & m &:& oxo3 \conc  LasR  \srewrites{0.25} R3   \\
(R_{5}) & m &:& R3  \srewrites{400} oxo3 \conc  LasR  &{}& (R_{6}) & m &:& R3 \conc  LasORI  \srewrites{0.25} RO3 \\
(R_{7}) & m &:& RO3  \srewrites{10} R3 \conc LasORI  &{}& (R_{8}) & m &:& RO3  \srewrites{1200} RO3 \conc  LasR  \\
(R_{9}) & m &:& RO3  \srewrites{300} RO3 \conc  LasI  &{}& (R_{10}) & m &:& LasI  \srewrites{1}  \emptyseq \\
(R_{11}) & m &:& LasR  \srewrites{1}  \emptyseq &{}& (R_{12}) & m &:& oxo3  \srewrites{1}  \emptyseq\\
(R_{13}) & m &:& I \conc  oxo3  \srewrites{0.1} A \conc oxo3  &{}& & & &
\end{array}
\end{equation}

Rules $R_1$ and $R_2$ describe the production from the DNA of proteins LasR
and LasI, respectively. Rule $R_3$ describes the production of the autoinducer,
modelled as atom $oxo3$, performed by the LasI enzyme. Rules $R_4$ and $R_5$ describe the
complexation and decomplexation of the autoinducer and the LasR protein,
where the complex is denoted $R3$. Rules $R_7$, $R_8$, $R_9$ describe the binding of the
activated autoinducer (the $RO3$ complex) to the DNA and its influence in the
production of LasR and LasI.  Finally, rules
$R_{10}$, $R_{11}$, $R_{12}$ describe the degradation of proteins. In particular, rule $R_{12}$ models the degradation of the autoinducer, which can happen both inside and outside the bacterium. Bacteria are marked as ``active'' (atom $A$) when reaching a sufficient high level of auto-inducer inside the respective sector otherwise they are marked as ``inactive'' (atom $I$).

The following rules describe the ability of the
autoinducer to cross the membrane, in both directions inside the sectors (different kinetics are used to model the different bandwidths of the channels). An autoinducer inside the bacterium is modelled as a
non-positional element, while autoinducers outside have an associated spatial sector $S_i$ for $i=1, \ldots, 4$:

\begin{equation}
\begin{array}{lcrl}
&S_i &:& ( x \into oxo3 \conc X)^{m} \srewrites{30} oxo3 ( x \into X)^{m}\\
&S_i &:& oxo3 ( x \into X)^{m} \srewrites{1} ( x \into oxo3 \conc X)^{m}\\
&S_i &:& oxo3  \srewrites{1}  \emptyseq
\end{array}
\end{equation}

The diffusion of the autoinducer across the different sectors is described by the following rules:

\begin{equation}
\begin{array}{lcrl}
&\TOP & : & ( x \into oxo3 \conc X)^{S_i} ( y \into Y)^{S_j}  \srewrites{k_{ij}} ( x \into  X)^{S_i} (y \into oxo3 \conc Y)^{S_j}  \\
\end{array}
\end{equation}

where $k_{ij}$ is related to the sectors channel capacities. In our experiments the values of $k_{ij}$ are shown in Table \ref{kij} where $k_{ij}=k_{ji}$.

\begin{table}
  \centering
  \begin{tabular}{|c||c|c|c|c|}
    \hline
    $i \backslash j$ & 1 & 2 & 3 & 4 \\
\hline
\hline
    1 & ND & 0.0 & 0.0 & 0.1 \\
\hline
    2 & 0.0 & ND & 0.2 & 0.3 \\
\hline
    3 & 0.0 & 0.2 & ND & 0.3 \\
\hline
    4 & 0.1 & 0.3 & 0.3 & ND \\
    \hline
  \end{tabular}
  \caption{Values for the kinetic rates $k_{ij}$}\label{kij}
\end{table}

The simulations were carried out using the following initial term $T$ (see Figure \ref{FIG_QS}):

\begin{equation}
\begin{array}{lclclcl}
T_1 &=& ( \into  \conc 10*( \into I \conc LasORI)^{m})^{S_{1}} &{}& T_2 &=& ( \into  \conc 15*( \into I \conc LasORI)^{m})^{S_{2}}\\
T_3 &=&  ( \into  \conc 15*( \into I \conc LasORI)^{m})^{S_{3}} &{}& T_4 &=& ( \into  \conc 8*( \into I \conc LasORI)^{m})^{S_{4}}\\
T &=& T_1 \conc T_2 \conc T_3 \conc T_4 & & & & 
\end{array}
\end{equation}

The model above can be easily enriched by adding rules describing also the movement of the bacteria between the different sectors and bacteria replication.

\end{appendix}

\end{document}